\documentclass[aps,pra,twocolumn,preprintnumbers,showpacs,superscriptaddress,amsmath,amssymb]{revtex4}
\usepackage{dcolumn}
\usepackage{bm}
\usepackage{amssymb}
\usepackage[german, english]{babel}
\usepackage{graphicx}
\usepackage{color}
\usepackage[letterpaper,total={7in,9.5in},top=0.75in,left=0.75in]{geometry}

\begin{document}
\title{Bose-Einstein condensation of $^{86}$Sr}

\author{Simon Stellmer}
 \affiliation{Institut f\"ur Quantenoptik und Quanteninformation (IQOQI),
\"Osterreichische Akademie der Wissenschaften, 6020 Innsbruck,
Austria}
\affiliation{Institut f\"ur Experimentalphysik und
Zentrum f\"ur Quantenphysik, Universit\"at Innsbruck,
6020 Innsbruck, Austria}
\author{{Meng Khoon} Tey}
\affiliation{Institut f\"ur Quantenoptik und Quanteninformation (IQOQI),
\"Osterreichische Akademie der Wissenschaften, 6020 Innsbruck,
Austria}
\author{Rudolf Grimm}
 \affiliation{Institut f\"ur Quantenoptik und Quanteninformation (IQOQI),
\"Osterreichische Akademie der Wissenschaften, 6020 Innsbruck,
Austria}
\affiliation{Institut f\"ur Experimentalphysik und
Zentrum f\"ur Quantenphysik, Universit\"at Innsbruck,
6020 Innsbruck, Austria}
\author{Florian Schreck}
\affiliation{Institut f\"ur Quantenoptik und Quanteninformation (IQOQI),
\"Osterreichische Akademie der Wissenschaften, 6020 Innsbruck,
Austria}

\date{\today}


\begin{abstract}
We report on the attainment of Bose-Einstein condensation of $^{86}$Sr. This isotope has a scattering length of about +800\,$a_0$ and thus suffers from fast three-body losses. To avoid detrimental atom loss, evaporative cooling is performed at low densities around $3\times 10^{12}$\,cm$^{-3}$ in a large volume optical dipole trap. We obtain almost pure condensates of $5\times10^3$ atoms.
\end{abstract}

\maketitle

Quantum degenerate gases of atoms with two valence electrons are an exciting field of research. The electronic structure of these atoms is the basis for applications like ultraprecise optical clocks \cite{Ido2003rfs, Ludlow2008slc, Lemonde2009olc,Sterr2004toc,Lemke2009sol,Hong2005oot,Kohno2009odo}. Quantum degenerate samples open new possibilities such as the study of unique quantum many-body phenomena \cite{Hermele2009mio,Cazalilla2009ugo,Gorshkov2010tos,Gerbier2010gff} and novel schemes of quantum computation \cite{Daley2008qcw,Gorshkov2009aem,Reichenbach2009cns,Stock2008eog}. They are also an ideal starting point for the creation of molecules of two-electron atoms \cite{Ciurylo2004pso, Koch2008pfc} and molecules of two-electron atoms with alkali atoms \cite{Zuchowski2010urm,Guerout2010tgs}, which have applications in precision measurement \cite{Kotochigova2009pfa} or quantum simulations with spin-dependent anisotropic long range interactions \cite{Micheli2006atf}. The first two-electron atom cooled to quantum degeneracy was ytterbium \cite{Takasu2003ssb,Fukuhara2007bec,Fukuhara2009aof}, followed by calcium \cite{Kraft2009bec} and strontium last year. Quantum degenerate samples of bosonic $^{84}$Sr and $^{88}$Sr \cite{Stellmer2009bec,deEscobar2009bec,Mickelson2010bec}, as well as of fermionic $^{87}$Sr \cite{DeSalvo2010dfg,Tey2010dgb} have been created, but Bose-Einstein condensation (BEC) of $^{86}$Sr has so far been elusive \cite{Ferrari2006cos}.

The success of evaporative cooling with the goal to reach quantum degeneracy depends largely on the ratio of elastic to inelastic collisions of the atomic species used \cite{Ketterle1996eco}. The four isotopes of strontium differ significantly in this respect. Despite its low natural abundance of 0.56\%, $^{84}$Sr was the first to be Bose condensed as it offers ideal scattering properties \cite{Stellmer2009bec,deEscobar2009bec}. $^{88}$Sr, which has a scattering length close to zero, and spin-polarized fermionic $^{87}$Sr barely collide at ultralow temperatures. They could only be cooled to quantum degeneracy using mixtures with other isotopes or spin states \cite{Mickelson2010bec,DeSalvo2010dfg,Tey2010dgb}. $^{86}$Sr poses the opposite challenge: its large scattering length of +800\,$a_0$ \cite{Stein2010tss} leads to a large three-body loss rate constant \cite{Ferrari2006cos}. Previously reported conditions of $^{86}$Sr in a dipole trap led to an elastic to inelastic collision ratio that was insufficient for evaporative cooling \cite{Ferrari2006cos}. Magnetic tuning of the scattering properties, as was an essential ingredient to circumvent similar difficulties on the way to BEC of cesium \cite{Weber2003bec}, is not available in alkaline-earth elements. Another approach to increase the ratio of elastic two-body collisions and inelastic three-body collisions is to lower the density. This approach was important for BEC of $^{40}$Ca, which is also an isotope with large scattering length \cite{Kraft2009bec}, as well as for BEC of cesium.

In this Rapid Communication, we show that it is indeed possible to create $^{86}$Sr BECs by performing evaporation at comparatively low densities around $3\times10^{12}$\,cm$^{-3}$. To obtain sizeable atom numbers at these low densities, we implement a large volume optical dipole trap.

Our experimental procedure initially follows the one used for Bose condensation of $^{84}$Sr \cite{Stellmer2009bec}. Zeeman slowed atoms are captured and cooled by a ``blue'' magneto-optical trap (MOT) operating on the $5s^{2}\,{^1S_0}-5s5p\,{^1P_1}$ transition at a wavelength of 461\,nm. A weak leak of the excited state of the cooling cycle continuously populates the $5s5p\,{^3P_2}$ metastable state. Weak-field seeking atoms in this state can be trapped in a magnetic trap formed by the quadrupole magnetic field used for the MOT. Metastable state atoms are accumulated in the magnetic trap for 1\,s. Further cooling and density increase is achieved by operating a ``red MOT'' on the 7.4\,kHz linewidth $^1S_0-{^3P_1}$ intercombination line at 689\,nm. To increase the capture velocity of the red MOT, we frequency modulate the light, producing sidebands, which cover a detuning range from a few ten kHz to a few MHz to the red of the transition. To load the red MOT, the metastable-state atoms in the magnetic trap are optically pumped to the $^1S_0$ ground state using the $5s5p\,{^3P_2}-5s5d\,{^3D_2}$ transition at 497\,nm. After
loading, the MOT is compressed by reducing the MOT beam
intensity and ramping off the frequency modulation, resulting
in a colder and denser sample. At this point the MOT contains $2\times 10^7$ atoms at a temperature of $1\,\mu$K.

The following evaporative cooling stage significantly differs from our previous experiments on $^{84}$Sr. The atoms are transferred into a crossed-beam optical dipole trap, which is much larger than the one used before. The trapping geometry consists of a horizontal and a nearly vertical beam, derived from a broadband ytterbium fiber laser operating at 1075\,nm. The horizontal beam has an elliptic beam shape with a horizontal waist of 300\,$\mu$m and a vertical waist of 33\,$\mu$m. The vertical beam is circular with a waist of 290\,$\mu$m and is used to provide additional confinement along the weak direction of the horizontal beam towards the end of evaporation. Initially, only the horizontal dipole trap beam is used and set to a power of 2\,W. The resulting trap is oblate with horizontal trap oscillation frequencies of 3\,Hz and 30\,Hz, a vertical trap oscillation frequency of 260\,Hz and a potential depth of 3.7\,$\mu$K, taking into account gravitational sagging. This oblate trap geometry combines the requirement of strong enough confinement against gravity in the vertical direction with the requirement of a large trap volume.

For optimum loading of the dipole trap, we adjust the
intensity and detuning of the red MOT beams before switching
them off. After 1\,s of plain evaporation, $2.5\times 10^6$ atoms at a temperature of 500\,nK remain in the trap. The peak density is $3\times 10^{12}$\,cm$^{-3}$ and the peak phase-space density is 0.05. Even at the temperature of 500\,nK, unitary limitation leads to a reduction of the thermally averaged elastic scattering cross section by a factor of two compared to the zero temperature value. The elastic collision rate is 380\,s$^{-1}$.

After the plain evaporation stage, forced evaporation is performed over 4.8\,s. The power of the horizontal dipole trap beam is reduced nearly exponentially to 740\,mW, with a longer time constant during the last 1.5\,s. During the first 2.3\,s of forced evaporation, the vertical dipole trap beam is increased to a power of 0.5\,W, at which it stays for the rest of the evaporation sequence. The vertical beam increases the confinement along the weak direction to 5\,Hz and does not provide any confinement along the vertical direction. The potential depth of the vertical beam in the radial directions is 200\,nK. During evaporation atoms escape the trap mainly downwards by leaving the trap over the potential barrier formed by the horizontal beam and gravity. It is beneficial for evaporation that the vertical trap oscillation frequency remains higher than the elastic collision rate, so that high energy atoms produced by collisions can quickly move out of the trap. During evaporation the peak density slightly drops to $10^{12}$\,cm$^{-3}$ and the elastic collision rate drops to 70\,s$^{-1}$ before condensation. At this point the temperature is $\sim$30\,nK and the thermally averaged scattering cross section is very close to the zero-temperature value.

\begin{figure}
\includegraphics[width=\columnwidth]{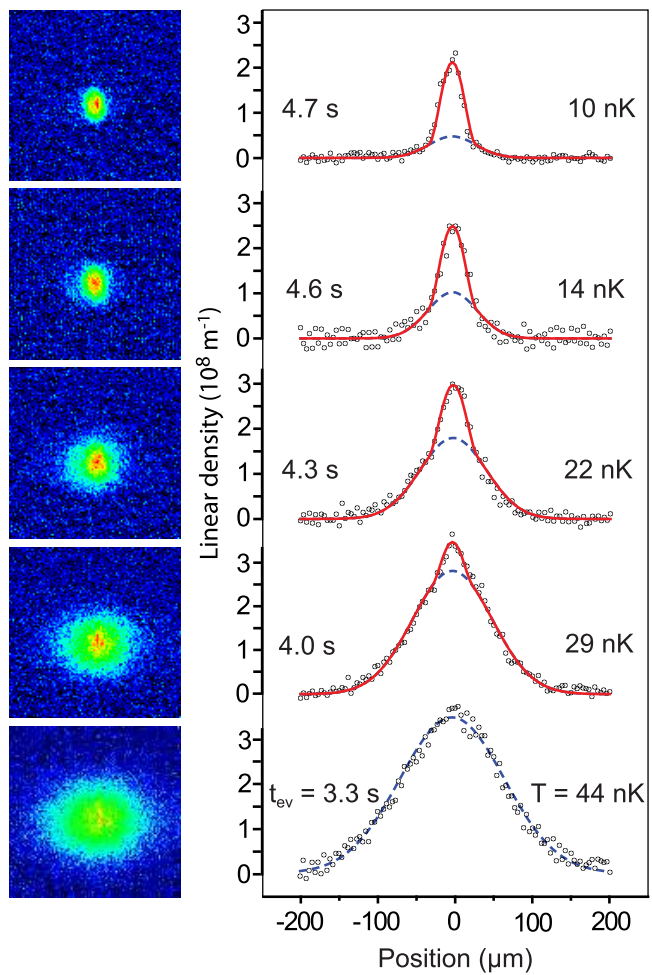}
\caption{\label{fig:Fig1} (Color online) Absorption images and integrated density profiles showing the BEC phase transition for different times $t_{\rm  ev}$ of the evaporative cooling ramp. The images are taken in the horizontal direction at 45$^{\circ}$ to the horizontal trap axes, 25\,ms after release from the trap. The field of view is 400\,$\mu$m by 400\,$\mu$m. The solid line represents a fit with a bimodal distribution, while the dashed line shows the Gaussian-shaped thermal part, from which the given temperatures are derived.}
\end{figure}

The phase transition from a thermal cloud to a BEC becomes evident in the appearance of a bimodal distribution, as clearly visible in time-of-flight absorption images and in the corresponding linear density profiles shown in Fig.~\ref{fig:Fig1}. At higher temperature the distribution is thermal, exhibiting a Gaussian shape. Cooling below the critical temperature $T_c$ leads to the appearance of an additional, narrower and denser, elliptically shaped
component: the BEC. The phase transition occurs after 4\,s of evaporative cooling, when the power of the horizontal beam is 820\,mW. At this point, the horizontal trap oscillation frequencies are 5\,Hz and 18\,Hz, the vertical trap oscillation frequency is $f_{\rm vert}=115$\,Hz and the trap depth is about 110\,nK, taking into account gravitational sagging. Note that despite the large scattering length, we are not in the collisionally hydrodynamic regime since the elastic scattering rate is only one tenth of $2\pi f_{\rm vert}$ \cite{Kagan1997eoa}.

\begin{figure}
\includegraphics[width=\columnwidth]{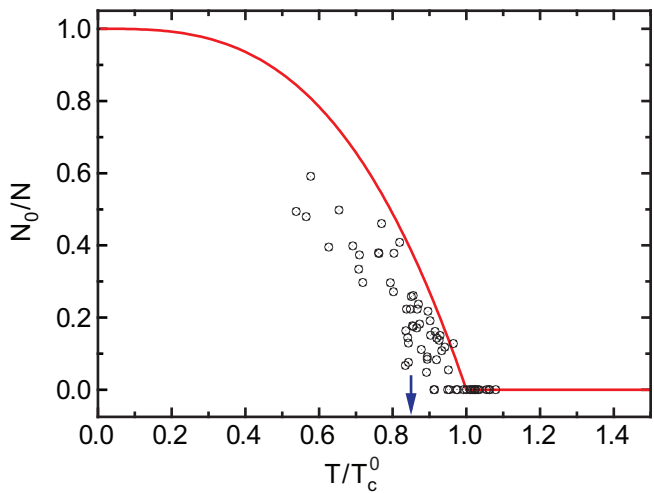}
\caption{\label{fig:Fig2} (Color online) The condensate fraction $N_0/N$ as a function of the temperature scaled by the critical temperature of an ideal gas in the thermodynamic limit $T_c^0$. The line shows the behavior expected for an ideal gas. The arrow indicates the expected transition temperature $T_c=0.85\,T_c^0$ when interactions and minor contributions from finite size effects are taken into account.}
\end{figure}

In order to analyze the phase-transition precisely, we fit bimodal distributions to the time-of-flight absorption images. The bimodal distributions consist of a Gaussian for the thermal part of the cloud and an inverted paraboloid integrated along one direction for the BEC. We extract the amount of condensed atoms $N_0$, the total amount of atoms $N$ and the temperature $T$ of the sample from the fits \cite{EndnoteFitDetails}. The condensate fraction $N_0/N$ as a function of temperature, scaled by the transition temperature of a non-interacting gas in the thermodynamic limit $T_c^0$ is shown in Fig.~\ref{fig:Fig2}. From this data we extract a critical temperature of $T_c=29(2)\,$nK$=0.90(5)\,T_c^0$. The atom number at the onset of BEC is $3.5\times 10^4$. Interaction effects reduce the critical temperature $T_c$ by 14\% compared to $T^0_c$, whereas finite-size effects add only a minor reduction of 1\%. These effects give a theoretically expected transition temperature of $T_c=0.85\,T_c^0$ \cite{Giorgini1996cfa}, which is within the uncertainty of our data.

\begin{figure}
\includegraphics[width=70mm]{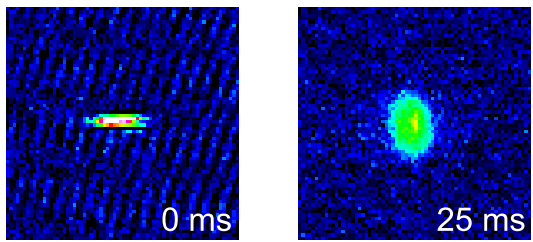}
\caption{\label{fig:Fig3} (Color online) Inversion of aspect ratio during the expansion of a pure BEC. The images (fields of view  $300\,\mu$m $\times$ $300\,\mu$m) are taken along the horizontal direction and show the BEC in the trap (left) and after 25\,ms free expansion (right).}
\end{figure}

After 4.8\,s of evaporation, a nearly pure BEC of $5\times10^3$ atoms is produced in a trap with oscillation frequencies of 5\,Hz and 16\,Hz in the horizontal plane and 70\,Hz in the vertical direction. The peak-density of the BEC is $3\times10^{12}$\,cm$^{-3}$ and the chemical potential is 7\,nK.  The BEC has an oblate shape with calculated {\it in-situ} Thomas-Fermi radii of 40\,$\mu$m and 12\,$\mu$m in the horizontal plane and 3\,$\mu$m in the vertical direction. After release from the trap, the BEC undergoes a mean-field driven expansion, with the strongest expansion in the vertical direction, resulting in an inversion of ellipticity as shown on the absorption images in Fig.~\ref{fig:Fig3}.

We extract an upper bound for the three-body loss rate constant from the lifetime measurement of a nearly pure BEC. The initial atom loss rate per atom from the BEC is $N_0^{-1} dN_0/dt=3(1)$\,s$^{-1}$. This value is determined from the initial slope of an exponential fit to 500\,ms of decay data. Assuming only three-body loss, the loss of atoms is described by $dN_0/dt=-(1/6)K_3\int n_0^3 dV$, where $n_0$ is the density distribution of the BEC and the factor $1/6$ takes into account the difference of three-body correlations of a thermal gas and a BEC. The resulting upper bound for the three-body loss rate constant is $K_3=6(3)\times10^{-24}$\,cm$^{6}$/s. This value is about six times higher than the value measured by Ferrari {\it et al.} \cite{Ferrari2006cos}. Moreover it is an order of magnitude higher than the maximally expected loss rate constant, which is $K_3=210 \hbar a^4/m=5\times10^{-25}$\,cm$^{6}$/s, assuming that three atoms are lost during each three-body loss event \cite{Fedichev1996tbr,Bedaque2000tbr}. Here $a$ is the scattering length and $m$ is the mass of $^{86}$Sr. Several explanations are possible for the unusually large measured value. More than three atoms might be lost per three-body loss event resulting from secondary collisions, possibly augmented by an enhanced atom-dimer scattering cross section \cite{Zaccanti2009ooa,Pollack2009uit}. Evaporation of atoms heated by technical noise of the dipole trap is an example for an explanation other than three-body loss. Further studies are needed to resolve this issue.

In conclusion, we have produced a Bose-Einstein condensate of $^{86}$Sr containing $5\times 10^3$ atoms. With this achievement all stable isotopes of strontium have been cooled to quantum degeneracy. The large scattering length of $^{86}$Sr leads to a very large three-body loss rate coefficient, which poses a challenge for evaporative cooling. We have shown that performing evaporation at low density in a large volume trap is a possible way to overcome this problem. The BEC of $^{86}$Sr enriches the possibilities opened up by strontium quantum gases. The large scattering length of $^{86}$Sr originates from a weakly bound state in the molecular potential. This state will influence the properties of optical Feshbach resonances \cite{Ciurylo2005oto,Enomoto2008ofr}. The BEC of $^{86}$Sr increases the options for quantum degenerate mixtures of different Sr isotopes or of mixtures of Sr with other elements. This wider choice can be important, for example in the search for magnetic Feshbach resonance in Sr-alkali mixtures suitable for molecule creation \cite{Zuchowski2010urm}.

We thank P. S. Julienne for fruitful discussions. We gratefully acknowledge support by the Austrian Ministry of Science and Research (BMWF) and the Austrian Science Fund (FWF) in the form of a START grant under project number Y507-N20. We also gratefully acknowledge support by the European Commission under project number 250072 iSENSE.

\bibliographystyle{apsrev}

\begin{thebibliography}{42}
\bibitem[{End()}]{EndnoteFitDetails}
\bibinfo{note}{To determine if a BEC is present, we also fit single Gaussian
  distributions to the images. If the fit residue of such a fit is comparable
  to that of a bimodal fit, we consider the distribution as purely thermal and
  extract temperature and atom number from the single Gaussian fit, otherwise
  from the bimodal fit. The temperature is determined from the size of the
  Gaussian component in the vertical direction, in which the cloud has expanded
  far enough beyond its initial size to show the momentum distribution.}
\end{thebibliography}

\end{document}